\documentclass[10pt, twocolumn]{extarticle}
\usepackage{graphicx}
\usepackage{hyperref}
\usepackage{amssymb}
\usepackage{amsmath}
\usepackage{amsfonts}
\usepackage{parskip}
\usepackage{titling}
\usepackage[table]{xcolor}
\usepackage[square,sort&compress,comma,numbers]{natbib}
\usepackage{multirow}
\usepackage{empheq}

\usepackage{titlesec}
\usepackage{caption}
\titleformat{\section}{\large\bf}{\thesection}{1em}{}
\titleformat{\subsection}{\bf}{\thesubsection}{1em}{}
\titleformat{\subsubsection}{\it}{\thesubsubsection}{1em}{}

\pretitle{\Large\bf\vskip 0.3em} \posttitle{\par\vskip 0.6em} \preauthor{\large} \postauthor{\par\vskip -0.0em}
\predate{\large} \postdate{\par\vskip 0.1em \rule{\linewidth}{0.5pt}} 

\newcommand{\beq}{\begin{equation}}
\newcommand{\eeq}{\end{equation}}
\newcommand{\bea}{\begin{eqnarray}}
\newcommand{\eea}{\end{eqnarray}}

\newcommand{\comment}[1]{}
\renewcommand{\d}{{\rm d}}

\begin{document}
\captionsetup{font=small}

\title{Accurate Self-Configuration of Rectangular Multiport Interferometers}
\author{Ryan Hamerly$^{1,2}$, Saumil Bandyopadhyay$^1$, and Dirk Englund$^1$}
\date{August 5, 2022}


\maketitle



\begin{flushleft}
\small
$^{1}$      \textit{Research Laboratory of Electronics, MIT, 50 Vassar Street, Cambridge, MA 02139, USA} \\
$^{2}$      \textit{NTT Research Inc., Physics and Informatics Laboratories, 940 Stewart Drive, Sunnyvale, CA 94085, USA}
\end{flushleft}

{\bf\small Abstract---Multiport interferometers based on integrated beamsplitter meshes are widely used in photonic technologies.  While the rectangular mesh is favored for its compactness and uniformity, its geometry resists conventional self-configuration approaches, which are essential to programming large meshes in the presence of fabrication error.  Here, we present a new configuration algorithm, related to the $2\times 2$ block decomposition of a unitary matrix, that overcomes this limitation.  Our proposed algorithm is robust to errors, requires no prior knowledge of the process variations, and relies only on external sources and detectors.  We show that self-configuration using this technique reduces the effect of fabrication errors by the same quadratic factor observed in triangular meshes.  This relaxes a significant limit to the size of multiport interferometers, removing a major roadblock to the scaling of optical quantum and machine-learning hardware.}

\rule{\linewidth}{0.5pt}

\section{Introduction}

\begin{figure}[b!]
\begin{center}
\includegraphics[width=1.0\columnwidth]{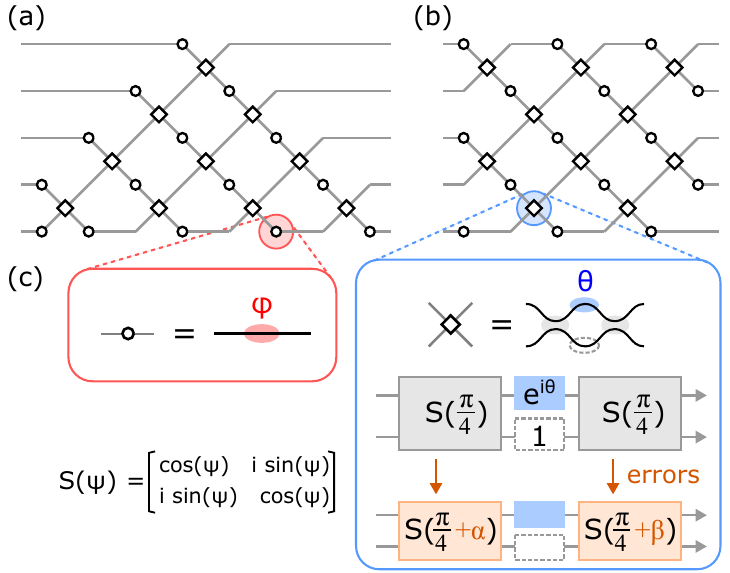}
\caption{(a) \textsc{Reck} and (b) \textsc{Clements} $5\times 5$ MZI meshes.  (c) Tunable building blocks: MZI and phase shifter.  Effect of errors on MZI transfer function.}
\label{fig:f1}
\end{center}
\end{figure}

Large-scale programmable photonic circuits are the cornerstone of many emerging technologies, including quantum computing \cite{Carolan2015, Zhong2020}, machine learning acceleration \cite{Shen2017, Tait2017, Hamerly2019}, and microwave photonics \cite{Marpaung2013, Zhuang2015}.  One such circuit---the universal multiport interferometer, which functions as a linear optical input-output device with a programmable transfer matrix (Fig.~\ref{fig:f1})---is of special importance due to its generality and broad range of applications \cite{Harris2018}.  The most scalable designs involve meshes of Mach-Zehnder interferometers (MZIs): while the triangular \textsc{Reck} mesh (Fig.~\ref{fig:f1}(a)) was initially employed and is straightforward to configure \cite{Reck1994}, more recently work has shifted to the \textsc{Clements} rectangle (Fig.~\ref{fig:f1}(b)), which offers clear advantages of compactness, path-length uniformity, and reduced sensitivity to loss \cite{Clements2016}.  Much recent study has focused on scaling \cite{Harris2020, Ramey2020} and optimizing \cite{Burgwal2017, Mower2015, Pai2019, Pai2020} MZI meshes based on the \textsc{Clements} design.

A major challenge to scaling MZI meshes is the presence of component errors due to fabrication imperfections.  Errors cause each MZI transfer matrix to deviate from its programmed value (Fig.~\ref{fig:f1}(c)); since the overall circuit is a cascade of MZIs with $O(N)$ depth, the total error in the transfer matrix scales as $O(\sqrt{N})$, where $N$ is the circuit size (assuming uncorrelated errors).  At large mesh sizes, these errors (if left uncorrected) place unreasonable constraints on fabrication tolerances, ultimately limiting the scaling of multiport interferometers.  Error-correction techniques are therefore critical for large-scale programmable photonics.  Global optimization \cite{Burgwal2017, Mower2015, Pai2019} and in-situ training \cite{Hughes2018} are promising in principle, but computationally inefficient and the optimization result is device-specific.  Local per-MZI correction is also effective, but requires prior characterization of the device errors \cite{SaumilPaper, Kumar2021}.  A number of efficient ``self-configuring'' algorithms have been proposed \cite{Miller2013a, Miller2013b, Miller2017}, but these generally only work for triangular meshes and require large numbers of internal power monitors \cite{Grillanda2014}, a significant addition in hardware complexity.  Recently, we proposed an efficient error-correction algorithm that does not require internal detectors or accurate pre-characterization \cite{RyanPaper}.  However, this algorithm only works for triangular (i.e.\ \textsc{Reck}) meshes, which excludes the more efficient \textsc{Clements} design.

In this article, we present a  
self-configuring strategy that is naturally adapted to the \textsc{Clements} design and requires no additional hardware complexity beyond external sources and (coherent) detectors.  This algorithm proceeds by configuring the diagonals of the mesh, starting at the corners, in a manner that progressively zeroes out the elements of a target matrix through a sequence of Givens rotations.  Numerical experiments on imperfect MZI meshes show that our algorithm is stable and robust, and reduces errors by at least a quadratic factor (while sufficiently small errors are corrected exactly), consistent with the behavior observed for triangular meshes \cite{RyanPaper}.  This significantly relaxes the scaling constraints posed by 
 imperfections in realistic MZI meshes.  As an example, we consider the application to optical neural networks and show how this correction scheme provides a path to overcome the no-go results of Ref.~\cite{Fang2019}.

\begin{figure}[tbp]
\begin{center}
\includegraphics[width=1.00\columnwidth]{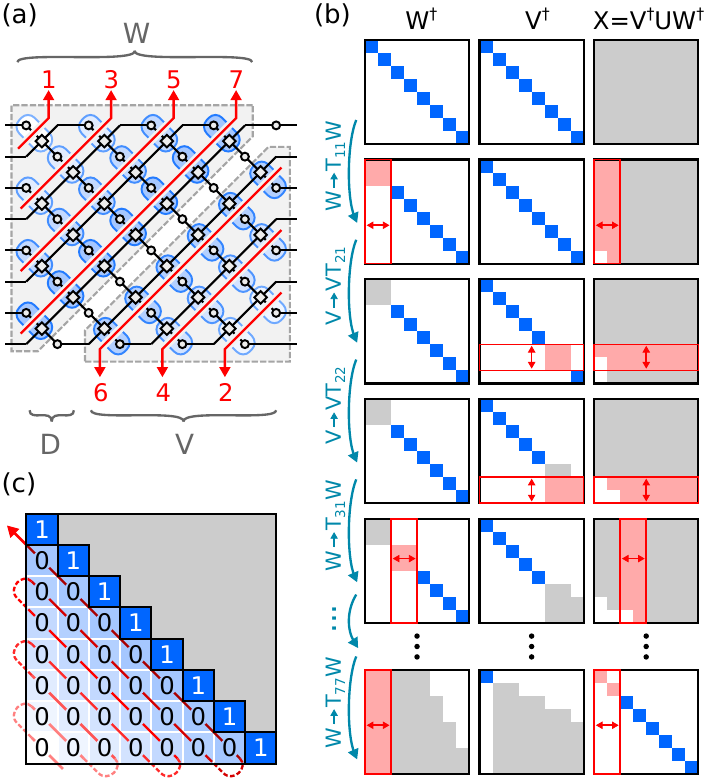}
\caption{(a) \textsc{Clements} mesh and order of MZI configuration.  (b) Building up the matrices $V, W$ in a sequence of Givens rotations to diagonalize $X = V^\dagger U W^\dagger$.  (c) Order of matrix elements zeroed by the procedure.}
\label{fig:f2}
\end{center}
\end{figure}

\section{Algorithm}
\label{sec:1}

Our procedure is based on the diagonalization of a unitary using $2\times 2$ Givens rotations \cite{Clements2016}, which we review here for clarity.  Following Fig.~\ref{fig:f2}(a), divide the mesh along the rising diagonal so that its transfer matrix becomes $U = V D W$, where $D$ is a phase screen and $V$ and $W$ represent the upper and lower triangles, given by:
\bea
	\!\!\!V & = & (T_{21} T_{22}) (T_{41} \ldots T_{44}) \ldots (T_{N-2,1}\ldots T_{N-2,N-2}) \nonumber \\
	\!\!\!W & = & (T_{N-1,N-1}\ldots T_{N-1,1})\ldots(T_{33}\ldots T_{31})(T_{11}) 
\eea
where $T_{mn}$ is the $2\times 2$ block unitary corresponding to the $n^{\rm th}$ crossing (MZI / phase-shift pair) of the $m^{\rm th}$ diagonal.  Following the order in the figure, we ``build up'' the matrices $(V, W)$ one block at a time while keeping track of $X = V^\dagger U W^\dagger$.  Fig.~\ref{fig:f2}(b) shows the first few steps of this process.  Updates to $W \rightarrow T_{mn}W$ right-multiply $X \rightarrow X T_{mn}^\dagger$, and the phases of $T_{mn}$ are chosen to zero an element in the lower-left corner of $X$.  Likewise, updates to $V \rightarrow V T_{mn}$ left-multiply $X \rightarrow T_{mn}^\dagger X$.  Following the order in Fig.~\ref{fig:f2}(c), this procedure zerores all elements in the lower-left triangle of $X$, enforcing diagonality.  The remaining phases are read off from the diagonal elements.

\begin{figure}[b!]
\begin{center}
\includegraphics[width=1.00\columnwidth]{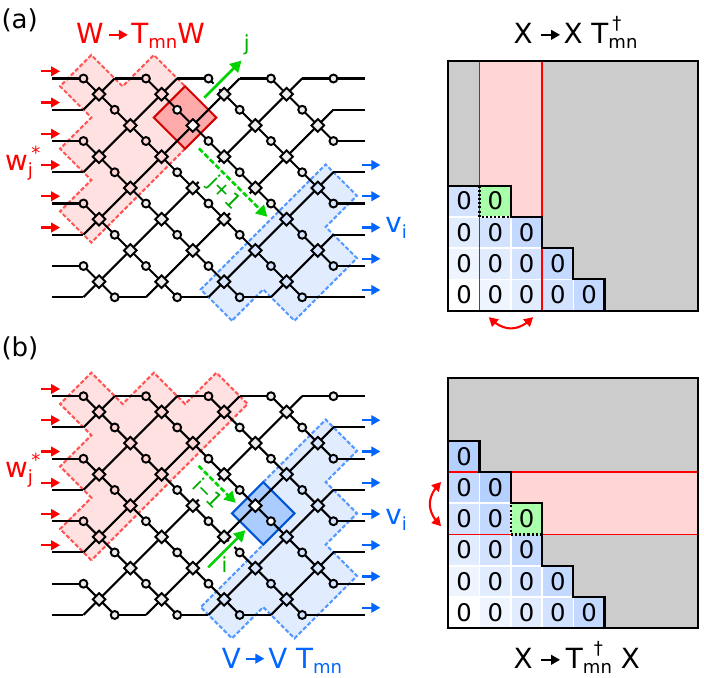}
\caption{Programming a physical \textsc{Clements} mesh in the presence of errors.  (a) Procedure to set phases for a crossing in $W$.  (b) Procedure for $V$.}
\label{fig:f3}
\end{center}
\end{figure}

While this procedure correctly sets the phases $(\theta, \phi)$ for an ideal \textsc{Clements} mesh, it does not work in the presence of errors because the relationship between $(\theta, \phi)$ and $T_{mn}$ also depends on the splitting angle imperfections $(\alpha, \beta)$ (Fig.~\ref{fig:f1}(c)), which are unknown.  Here, we describe the procedure for programming \textsc{Clements} in the presence of errors:

\begin{enumerate}
	\item Initialize the MZIs to approximate the cross state ($\theta = 0$).  An ideal cross state is not possible with errors, but an approximation will be fine.
	\item Configure the crossings in the order given by Fig.~\ref{fig:f2}(a).  For each crossing in $W$ [resp.\ $V$]:
	\begin{enumerate}
		\item Perform the Givens rotation $X \rightarrow X T_{mn}^\dagger$ [resp.\ $X \rightarrow T_{mn}^\dagger X$] that zeroes the next element of $X_{ij}$ in the sequence Fig.~\ref{fig:f2}(c) (note $(i, j) \neq (m, n)$).
		\item Update $W \rightarrow T_{mn}W$ [resp.\ $V \rightarrow V T_{mn}$].
		\item Send input $\vec{a}_{\rm in} = \vec{w}_j^*$ (the $j^{\rm th}$ column of $W^\dagger$) into the device, Fig.~\ref{fig:f3}.  The output $\vec{a}_{\rm out}(\theta, \phi)$ depends on the phases being configured.  Set $(\theta, \phi)$ to zero the inner product $\langle \vec{v}_i | \vec{a}_{\rm out}(\theta, \phi) \rangle$, where $\vec{v}_i$ is the $i^{\rm th}$ column of $V$.
	\end{enumerate}
	\item Finally, set the diagonal: for each $i$, inject $\vec{a}_{\rm in} = w_i^*$ and adjust $\phi$ to satisfy $\text{arg}(\langle \vec{v}_i | \vec{a}_{\rm out}(\phi) \rangle) = \text{arg}(X_{ii})$.
\end{enumerate}

Of the main loop (Step 2), parts (a-b) are just a restatement of the \textsc{Clements} factorization \cite{Clements2016}.  However this is performed merely to keep track of matrices $(V, W)$; we do not use the $(\theta, \phi)$ provided by the algorithm.  Instead, Step 2(c) uses physical measurements find the correct phases in the presence of hardware errors.  This step, which amounts to zeroing matrix element $X_{ij} = v_i^* U w_j^*$ of the physical hardware, is visualized in Fig.~\ref{fig:f3}:
\begin{enumerate}
	\item When configuring an $W$ (Fig.~\ref{fig:f3}(a)), we input $\vec{a}_{\rm in} = \vec{w}_j^*$ and program $(\theta, \phi)$ to direct all the light to intermediate output $j$ (solid arrow), zeroing the power that goes to $j+1$ (dashed arrow).  We do not have access to these intermediate outputs, but $j+1$ connects to the input $i$ of $V$; therefore $\langle \vec{v}_i | \vec{a}_{\rm out}\rangle$ is a valid proxy for this field and zeroing it correctly configures the red block to match $W$.
	\item When configuring $V$ (Fig.~\ref{fig:f3}(b)), we  want $\vec{v}_i$ to be the output from light at intermediate port $i$ (solid arrow); therefore, the output from light at port $i-1$ (dashed arrow) should be orthogonal to $\vec{v}_i$.  We excite this field by inputting $\vec{w}_j^*$; any errors in the unconfigured mesh will hit downstream inputs $k < i-1$, but will not pollute input $i$.  Therefore zeroing $\langle \vec{v}_i | \vec{a}_{\rm out} \rangle$ correctly configures the blue block of the mesh to match $V$, provided $W$ is properly configured.
\end{enumerate}

In both cases, the nulling signal $\langle \vec{v}_i | \vec{a}_{\rm out}\rangle$ arises from light transmitted down the dashed diagonals in Fig.~\ref{fig:f3}; a (near) cross-state initialization is needed to maintain a non-negligible transmitted power.  As we discuss later, this initialization is limited by random phase shifts in the MZIs, which can be fixed with a one-time calibration.  It is also relatively straightforward to extend our proposed scheme to triangular meshes, as described in Appendix~\ref{sec:s1}.

\begin{figure}[b!]
\begin{center}
\includegraphics[width=1.00\columnwidth]{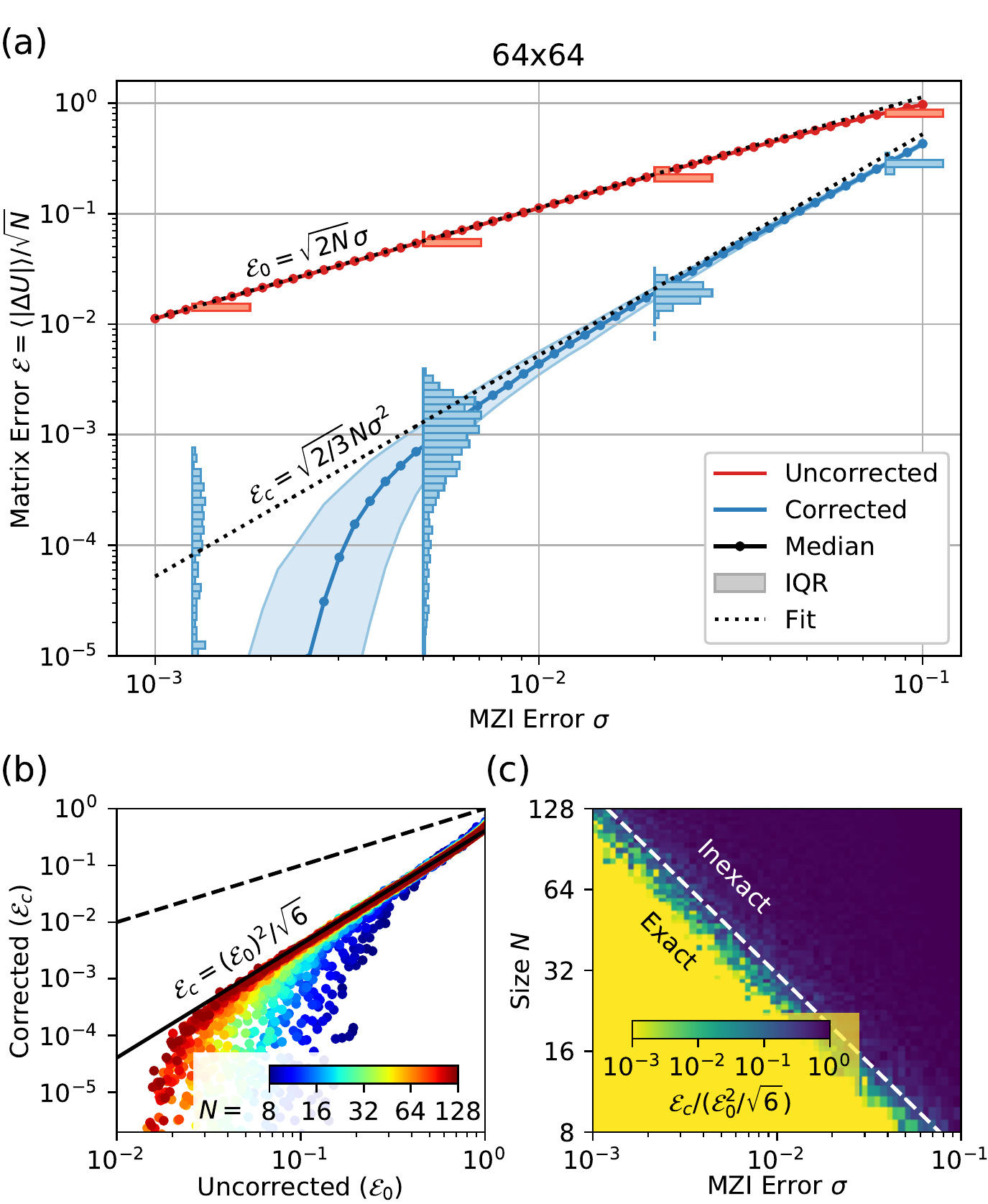}
\caption{(a) Corrected and uncorrected matrix error as a function of $\sigma$, $64\times 64$ \textsc{Clements} mesh.  (b) Scaling with mesh size, showing the quadratic suppression of errors due to correction.  (c) Boundary between the regimes of exact and inexact error correction.}
\label{fig:f4}
\end{center}
\end{figure}

\section{Results}

To test our algorithm, we performed numerical experiments on \textsc{Clements} meshes of size up to $128\times 128$.  The algorithms were implemented in \textsc{Python} and are available as part of the \textsc{Meshes} package \cite{Meshes}.  We consider an error model based on imperfections in splitter angles $\alpha, \beta$ (Fig.~\ref{fig:f1}(c)), as these are the dominant effect of fabrication error in MZI-based silicon photonic circuits.  For simplicity, consider the case of uncorrelated Gaussian errors so that the error magnitude can be characterized by a single variable $\sigma = \langle \alpha\rangle_{\rm rms} = \langle\beta\rangle_{\rm rms}$ (the case of correlated errors is treated in Appendix~\ref{sec:s2}; the qualitative results are the same because most correlations cancel out in ensemble averaging \cite{RyanPaper}).  Target matrices are sampled uniformly over the Haar measure \cite{Haar1933, Tung1985}.  As a figure of merit, we consider the normalized matrix error $\mathcal{E} = \langle\lVert \Delta U\rVert \rangle_{\rm rms}/\sqrt{N}$.  For unitary matrices, $\mathcal{E} \in [0, 2]$ corresponds to the average relative error of a given matrix element $U_{ij}$.

In the uncorrected case, each MZI introduces a mean error $\langle \lVert \Delta U\rVert \rangle_{\rm rms} = \sqrt{2}\sigma$.  These errors add in quadrature, leading to an overall normalized error $\mathcal{E}_0 = \sqrt{2N}\sigma$, which grows with mesh size.  This is understandable given that a circuit depth that grows as $O(N)$, with each layer contributing $O(\sigma)$ error and the layers adding in quadrature.  Fig.~\ref{fig:f4}(a) plots the error as a function of $\sigma$ for a $64\times 64$ mesh, both without error correction (red) and with our algorithm (blue).  The algorithm always improves the matrix fidelity, but there are two distinct regimes: for small $\sigma$, the error approaches machine precision, as errors can be corrected exactly.  On the contrary, for large $\sigma$, the corrected error asymptotes to a finite value:
\beq
	\mathcal{E}_c = \sqrt{2/3}\,N\sigma^2 = \frac{1}{\sqrt{6}}\mathcal{E}_0^2 \label{eq:ec}
\eeq
This form can be derived rigorously from the distribution of MZI splitting angles for Haar-sampled unitary matrices \cite{Russell2017, RyanPaper}, where errors arise solely from MZIs whose target splitting ratios cannot be realized in the imperfect hardware.  Since $\mathcal{E}_c \propto \mathcal{E}_0^2$, we can say that self-calibration leads to a {\it quadratic} suppression of errors: the smaller the initial error, the greater the benefit of error correction.

Fig.~\ref{fig:f4}(b-c) tests the scalability of the algorithm by varying the mesh size $N$.  Fig.~\ref{fig:f4}(b) shows that, as expected, as the mesh size grows, the corrected error asymptotes to the quadratic factor Eq.~(\ref{eq:ec}).  By plotting $\mathcal{E}_c/(\mathcal{E}_0^2/\sqrt{6})$, Fig.~\ref{fig:f4}(c) shows the boundary between the exactly correctable small-error regime ($\mathcal{E}_c = 0$) and the inexact large-error regime where Eq.~(\ref{eq:ec}) holds.  The regimes meet where the coverage $\text{cov}(N)$ of $U(N)$ dips below unity \cite{Russell2017}; we have previously shown that $\text{cov}(N) \sim e^{-N^3\sigma^2/3}$ \cite{RyanPaper}; therefore the boundary lies at roughly $N^3\sigma^2 = 3$ (dashed curve in Fig.~\ref{fig:f4}(c)).

\begin{figure}[tbp]
\begin{center}
\includegraphics[width=1.00\columnwidth]{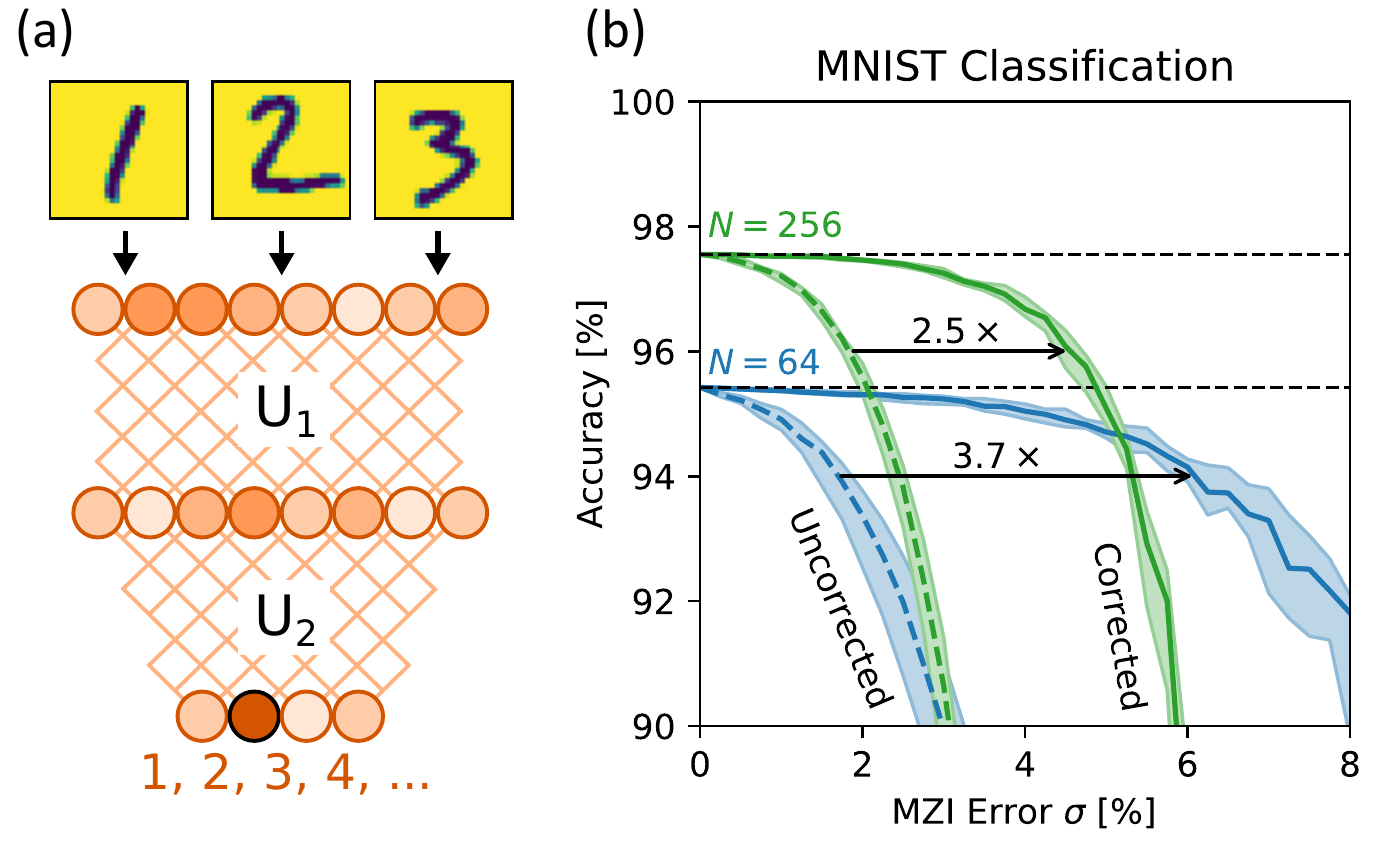}
\caption{(a) 2-layer DNN for MNIST classification where synaptic connections are represented by \textsc{Clements} meshes.  (b) Simulated classification accuracy as a function of network size $N$ and hardware MZI error $\sigma$.}
\label{fig:f5}
\end{center}
\end{figure}

To illustrate the benefits of this error reduction, we consider as an example deep neural network (DNN) inference on optical hardware.  DNNs process data in a sequence of layers, each consisting of (linear) synaptic connections and (nonlinear) neuron activation (Fig.~\ref{fig:f5}(a)).  One exciting possibility is to use photonics to accelerate this process: encode the input in optical amplitudes, use a programmable MZI mesh to implement the synaptic weights, and perform the activations with an all-optical (or electro-optic) nonlinearity \cite{Shen2017}.  A major challenge is that useful learning tasks require large mesh sizes ($N > 100$), which are particularly susceptible to fabrication error; a recent study showed that accurate DNN inference might require unrealistic process tolerances in the hardware \cite{Fang2019}.  This challenge has spurred investigations into alternative proposals, which have their own limitations \cite{Tait2017, Hamerly2019, Bernstein2020}.

Fig.~\ref{fig:f5}(b) illustrates the advantage of error correction in an MZI-mesh DNN accelerator.  Here, input images are preprocessed by a Fourier transform and fed into a two-layer DNN, with electro-optic neuron activations designed to approximate a complex modReLU \cite{Williamson2019, Zhang2021}.  Models with inner layer sizes $N = 64$ and $N=256$ were trained on the MNIST digit dataset \cite{LeCun1998} using the \textsc{Neurophox} package \cite{Neurophox}.  Details and code are provided in Appendix~\ref{sec:s3} and the Supplementary Material \cite{Supp}, respectively.  These pre-trained models were then simulated numerically on imperfect \textsc{Clements} meshes to calculate the classification accuracy.  Applying our correction algorithm increases the error tolerance of these DNNs by over $2\times$.  Note that directional couplers in silicon typically exhibit $\sigma \approx 2\%$ \cite{Mikkelsen2014}; in this regime, the uncorrected DNNs show significant degradation, while error correction restores them to their canonical accuracy.  Error correction may even allow the use of broad-band multi-mode interference couplers, which typically exhibit larger hardware errors.

\begin{figure}[t!]
\begin{center}
\includegraphics[width=1.00\columnwidth]{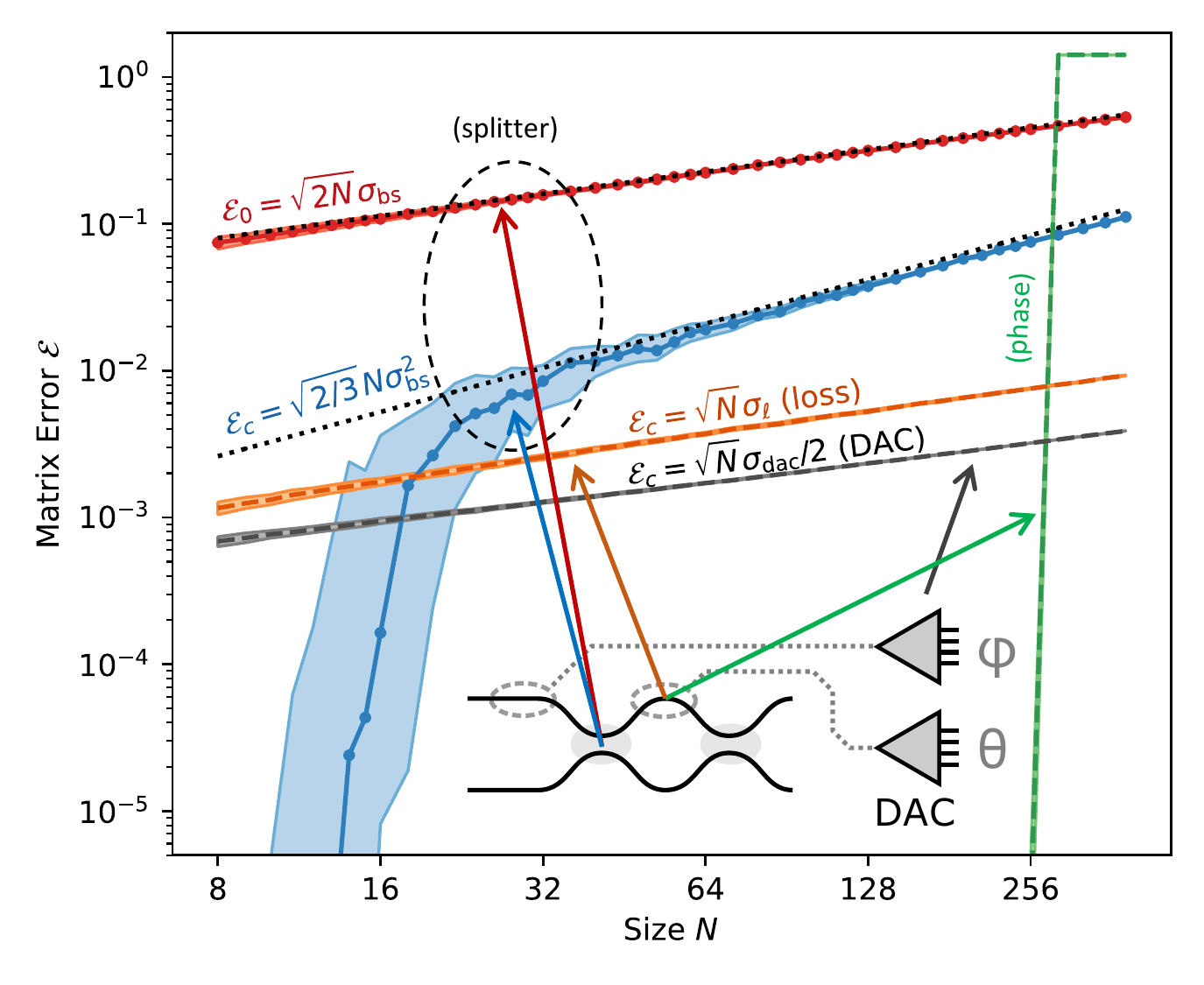}
\caption{Comparison of hardware error types: $\mathcal{E}$ plotted against mesh size.  Here $\sigma_{\rm bs} = 2\%$, $\sigma_{\rm ph} = 0.2$, $\sigma_\ell = 5\times 10^{-4}$, and $\sigma_{\rm dac} = 2\pi/\sqrt{12}\times 2^{-B}$ with $B = 12$ bits of precision.  See Appendix~\ref{sec:s4} for derivation.}
\label{fig:f6}
\end{center}
\end{figure}

\section{Discussion}

Beyond splitting-ratio errors, realistic meshes must contend with a range of additional imperfections, including undesired phase shifts, unbalanced losses, and limited programming precision.  We analyze these error sources in Appendix~\ref{sec:s4}, with the main results shown in Fig.~\ref{fig:f6}.  Phase-shift and loss errors introduce a random transmission coefficient $e^{i\psi - \ell/2}$ to each waveguide segment, where $\langle\psi\rangle_{\rm rms} = \sigma_{\rm ph}$, $\langle\ell\rangle_{\rm rms} = \sigma_{\ell}$.  If known, phase errors can be corrected exactly with a one-time calibration.  However, even without calibration, self-configuration still succeeds provided the errors are small enough, and we observe a sharp error-correction threshold of $\sqrt{N} \sigma_{\rm ph} \lesssim O(1)$.  This condition is related to the need to initialize the MZIs close enough to the near-cross state that a significant fraction of the light can propagate down the nulling diagonal (dashed green line in Fig.~\ref{fig:f3}) to give a non-negligible signal $\langle \vec{v}_i | \vec{a}_{\rm out} \rangle$.

Unlike phase errors, unbalanced losses and programming (DAC) errors are not correctable, as evidenced by the observed behavior $\mathcal{E}_c = \sqrt{N}\sigma_\ell$ and $\sqrt{N}\sigma_{\rm dac}/2$, which follows the same $\sqrt{N}$ scaling observed for uncorrected errors.  Loss imbalance is limited by statistical fluctuations in surface roughness; an analysis based on Ref.~\cite{Lacey1990} gives $\sigma_\ell \approx \pi^{-1}\alpha_{\rm wg} \sqrt{L_c L \log(4L/L_c)}$, where $\alpha_{\rm wg}$ is the waveguide loss, $L$ is the waveguide length, and the scatterer size $L_c = \lambda/n_0$ depends on wavelength and cladding refractive index.  For a $L = 200$~$\mu$m and $\alpha_{\rm wg} = 2$~dB/cm, one calculates $\sigma_\ell = 5\times 10^{-4}$, which is close to observed wafer-scale loss variations \cite{Wilmart2020}.  DAC error is set by the average truncation to $B$ bits of a signal in the range $[0, 2\pi]$: $\sigma_{\rm dac} = 2\pi/\sqrt{12} \times 2^{-B}$.  While neither loss or DAC error are correctable, Fig.~\ref{fig:f6} clearly shows that these effects are about an order of magnitude smaller than the typical splitter errors, even after correction.  These errors will only become relevant if the (correctable) unitary imperfections can be substantially reduced, e.g.\ with post-fabrication trimming \cite{Chen2017} or ``perfect'' MZI-doubled structures \cite{Miller2015}.

Our algorithm runs in $O(N^2)$ steps and requires only $O(N^3)$ FLOPs of computation.  This figure is at least an order of magnitude faster than in-situ or direct numerical optimization schemes (Appendix~\ref{sec:s5}).  While local error correction \cite{SaumilPaper} and parallelized progressive methods \cite{Pai2020} are faster still, these schemes respectively require calibration and internal detectors, a major challenge to their deployment on large-scale photonic meshes.


We have proposed a  
self-configuration technique for rectangular MZI meshes.  Our technique requires only external sources and (coherent) detectors and does not rely on an accurate characterization of device errors.  This method is based on the diagonalization of a unitary matrix by Givens rotations, with a specific set of measurements performed to ensure that the Givens rotations are properly implemented in the hardware.  For sufficiently small hardware errors, our approach leads to perfect realization of the target matrix.  For large errors, it achieves the same quadratic reduction $\mathcal{E} \rightarrow \mathcal{E}^2/\sqrt{6}$ observed for local correction algorithms \cite{SaumilPaper} and self-configuration on triangular meshes \cite{RyanPaper}.  As a target application, we considered optically accelerated DNNs and showed that the proposed technique increases their robustness to hardware error, particularly in the critical region around $\sigma \approx 2\%$ characteristic of directional couplers in silicon.

One open question is increasing the robustness of error correction to non-unitary errors (unbalanced losses), as many emerging photonic devices often have undesired state-dependent loss \cite{Li2019, Gill2009, Harris2014}.  In addition, extensions of the algorithm to more recently developed mesh geometries \cite{Lopez2019, Fldzhyan2020, Saygin2020, Tanomura2019, Polcari2018} may prove useful, as some geometries are less sensitive to hardware error and may be easier to scale up to large dimensions.

S.B.\ is supported by an NSF Graduate Research Fellowship under grant no.\ 1745302 and the Air Force Office of Scientific Research (AFOSR) under award number FA9550-20-1-0113.  D.E.\ acknowledges funding from AFOSR (no.\ FA9550-20-1-0113, FA9550-16-1-0391).  R.H., S.B., and D.E. are inventors on patent application No.~96/196,301 assigned to MIT and NTT Research that covers techniques to suppress component errors in interferometer meshes.

\appendix

\section{Triangular Meshes}
\label{sec:s1}

\begin{figure}[b!]
\begin{center}
\includegraphics[width=1.00\columnwidth]{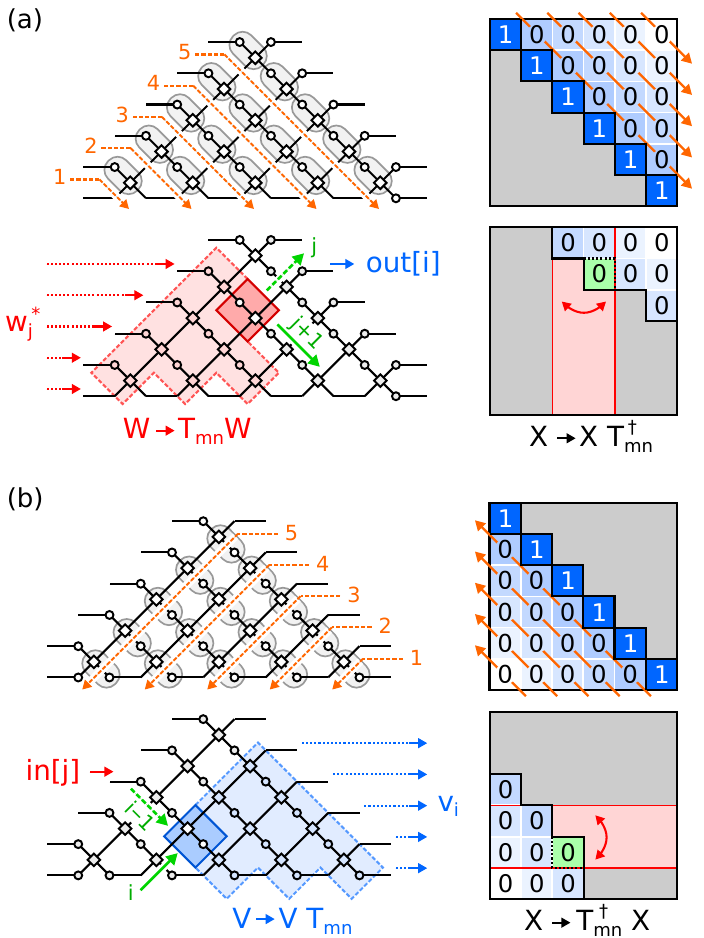}
\caption{Self-configuration procedure for the triangular \textsc{Reck} mesh.  The procedure depends on whether the mesh layout has (a) an output phase screen, or (b) an input phase screen.}
\label{fig:fs1}
\end{center}
\end{figure}

The key to self-configuring the \textsc{Clements} mesh \cite{Clements2016} was realizing that it could be divided into two triangles with a phase screen in the center: $U = V D W$.  By configuring one diagonal at a time, alternating between $V$ and $W$, we could zero all the elements of the target matrix $X = V^\dagger U W^\dagger$, thus realizing the desired unitary.  This procedure is simplified when configuring \textsc{Reck} \cite{Reck1994}, which can be expressed in terms of a single triangle:
\beq
	U = D \Bigl(\underbrace{\prod_{mn} T_{mn}}_{W}\Bigr)\ \ \text{or}\ \ \Bigl(\underbrace{\prod_{mn} T_{mn}}_{V}\Bigr) D
\eeq
The first case $U = D W$ corresponds to a mesh with an output phase screen (Fig.~\ref{fig:fs1}(a)), while the mesh for $U = V D$ has an input phase screen (Fig.~\ref{fig:fs1}(b)).  The algorithm in Sec.~\ref{sec:1} simplifies because we may substitute $V = I$ [resp.\ $W = I$].  This leads to two cases:

\begin{enumerate}
	\item For the first case (Fig.~\ref{fig:fs1}(a)), we work downstream from the leftmost MZI.  For example, we can work down the falling diagonals, with the order given in the figure.  Each step adds an MZI to $W \rightarrow T_{mn} W$, which updates the target matrix $X \rightarrow X T_{mn}^\dagger$ to zero an element in the upper triangle.  With input field $a_{\rm in} = w_j^*$ (the $j^{\rm th}$ column of $W^\dagger$), the parameters $(\theta, \phi)$ are set to zero the field at output $i$, where $(i, j)$ is the index of the element $X_{ij}$ being zeroed.
	\item For the second case (Fig.~\ref{fig:fs1}(b)), we work upstream from the rightmost MZI, which performs the update $V \rightarrow T_{mn} V$, $X \rightarrow T_{mn}^\dagger X$.  With light sent into the $i^{\rm th}$ input, ($\theta, \phi$) are chosen to zero the dot product between the output field and $v_i$ (the $i^{\rm th}$ column of $V$).
\end{enumerate}

After the crossings are configured, the phases of $D$ can be obtained by inspection.

The MZI order in Fig.~\ref{fig:fs1} is not unique.  Any order that preserves causality (the set of configured and unconfigured MZIs must be causally separated) will produce a valid self-configuration.

Like the configuration of the \textsc{Clements} rectangle, the procedure above is designed, at each time step, to properly set the configured MZIs in the red [resp.\ blue] block to realize $W$ [resp.\ $V$].  The procedure in Fig.~\ref{fig:f1}(a) is closely related to the Reversed Local Light Interference Method (RELLIM) \cite{Miller2017}.  However, there are two important differences.  First, RELLIM inputs columns of $U^\dagger$, while our scheme inputs columns of $W^\dagger$.  As the MZIs are configured, $W$ changes.  Second, RELLIM assumes the existence of internal power detectors after each MZI (or the ability to pre-calibrate the downstream mesh so that internal fields can be read off from the outputs).  By inputting a row of the target $W^\dagger$, light will exit the red block along only the $j^{\rm th}$ and $(j+1)^{\rm th}$ channels (provided all upstream MZIs are correctly set).  The former, which is set to zero by varying $(\theta, \phi)$, is read off directly from the $i^{\rm th}$ output. 

The procedures in Fig.~\ref{fig:fs1}(a-b) are reciprocal to each other and can also be related to the ``ratio method'' described in Ref.~\cite{RyanPaper}, since the process of zeroing matrix elements never depends on absolute amplitudes, only on their ratios.  Moreover, the ratio method is also robust in the presence of large errors and achieves the same quadratic error reduction.  However, the when configuring each MZI, the ratio method required a sweep of the upstream phase shifter in order to subtract the common amplitude $\vec{a}$, a procedure that is not necessary here.

\begin{figure}[tb]
\begin{center}
\includegraphics[width=1.00\columnwidth]{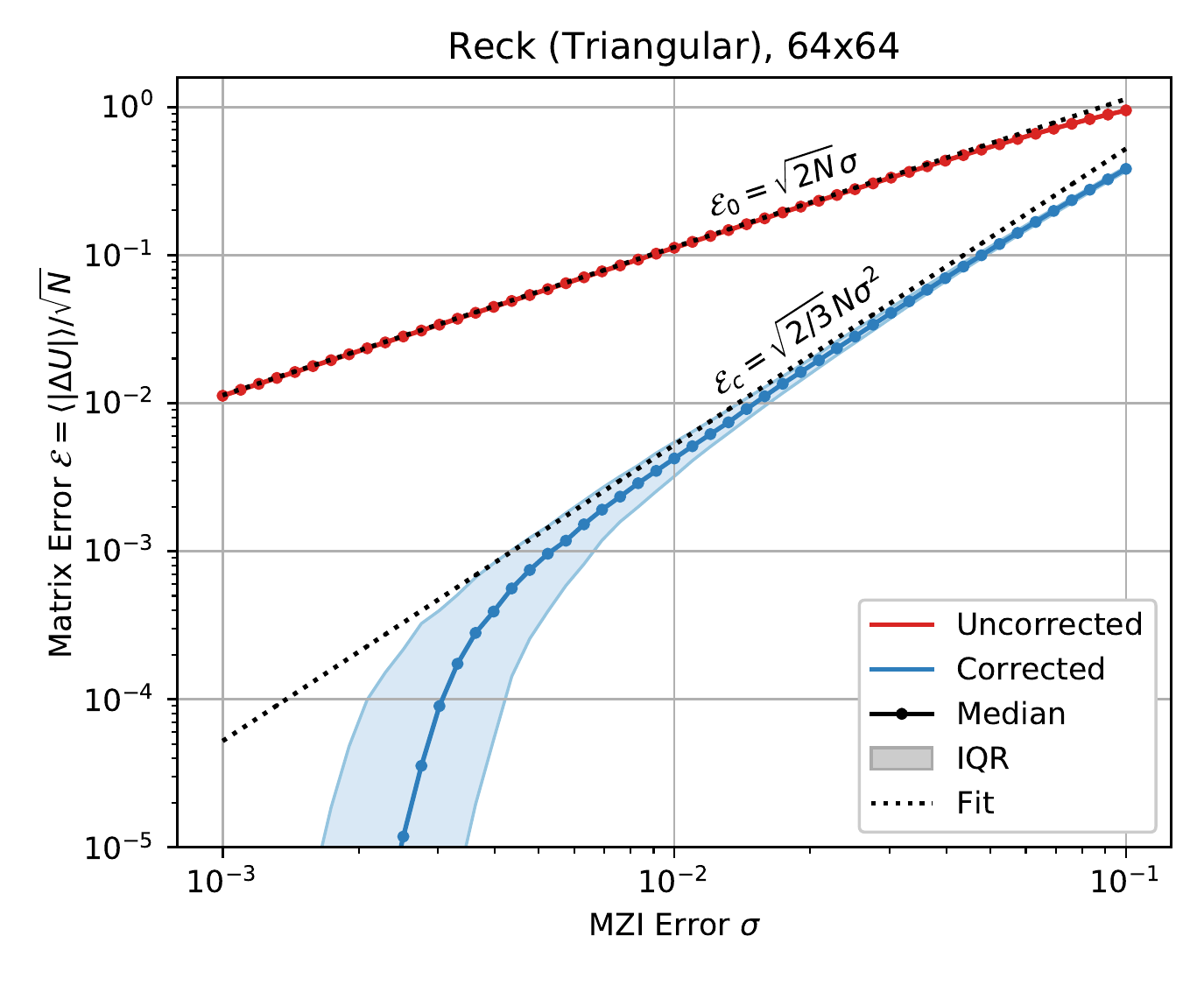}
\caption{Corrected and uncorrected matrix error as a function of MZI error $\sigma$, $64\times 64$ \textsc{Reck} mesh.}
\label{fig:fs2}
\end{center}
\end{figure}

Fig.~\ref{fig:fs2} plots the matrix error $\mathcal{E} = \lVert \Delta U \rVert / \sqrt{N}$ as a function of MZI error $\sigma$ for the \textsc{Reck} mesh.  The result is identical to the case of the \textsc{Clements} mesh, Fig.~\ref{fig:f4}, following the curves $\mathcal{E}_0 = \sqrt{2N}\sigma$, $\mathcal{E}_c = \sqrt{2/3}N\sigma^2$ predicted by theory \cite{RyanPaper}.  Over the Haar measure, the distribution of MZI splitting angles is the same for both mesh geometries up to a reordering of the MZIs \cite{Russell2017}; therefore, this correspondence is unsurprising.

The self-configuration procedures for \textsc{Reck} and \textsc{Clements} can be mapped to a general-purpose subroutine that self-configures any MZI mesh of the form $U = V D W$, provided that the geometry admits a matrix diagonalization by way of Givens rotations.  This algorithm has been implemented in \textsc{Python} with \textsc{Numba} extensions for numerical efficiency, and is available as part of the \textsc{Meshes} package \cite{Meshes}

\section{Correlated Errors}
\label{sec:s2}

\begin{figure}[tb]
\begin{center}
\includegraphics[width=1.00\columnwidth]{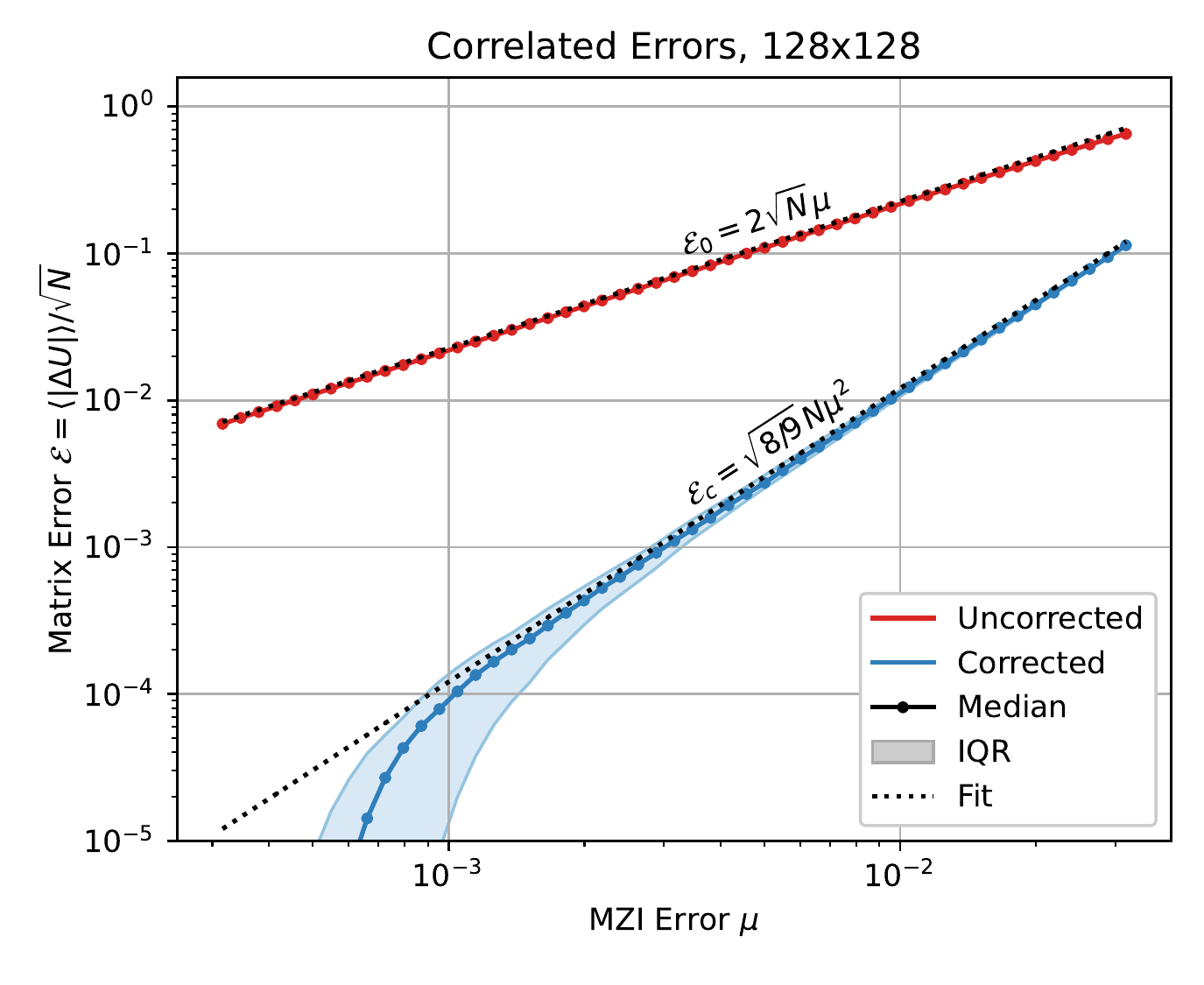}
\caption{Corrected and uncorrected matrix error as a function of correlated MZI error amplitude $\mu$, $128\times 128$ \textsc{Clements} mesh.}
\label{fig:fs3}
\end{center}
\end{figure}

In realistic MZI meshes, the splitter errors $\alpha_n, \beta_n$ will be strongly correlated, since the process variations that lead to errors (waveguide thickness and spacing, partial etch depth, slab height) all have correlation lengths much longer than the size of an MZI.  In general, the matrix error $\langle \lVert \Delta U \rVert^2 \rangle$ will depend both on the error amplitudes $(\langle \alpha_n^2 \rangle, \langle \beta_n^2 \rangle)$ as well as their correlations ($\langle \alpha_m\alpha_n\rangle$, etc.).  For an individual matrix $U$, the dependence on correlations can be significant.  However, we have previously shown that, in an {\it ensemble average} of matrices uniformly sampled over the Haar measure, this dependence becomes very small for most inter-MZI correlations because of the random phase shifts between pairs of MZIs \cite[Appendix~A]{RyanPaper}.  Only intra-MZI correlations $\langle \alpha_n \beta_n \rangle$ have a significant effect on the ensemble-averaged matrix error.

Consider the extreme case of full correlation $\alpha_n = \beta_n = \mu$.  This case is realized, for instance, when the dominant error source arises from operating the mesh away from the coupler design wavelength.  In Ref.~\cite{RyanPaper}, the coverage, uncorrected error, and corrected error are calculated to be:
\bea
	\text{cov}(N) & = & e^{-(2/3)N^3 \mu^2} \\
	\mathcal{E}_0 & = & 2\sqrt{N}\mu \\
	\mathcal{E}_c & = & \sqrt{8/9}N\mu^2
\eea
Fig.~\ref{fig:fs3} plots $\mathcal{E}_0$ and $\mathcal{E}_c$ against the error amplitude $\mu$ for a $128\times 128$ \textsc{Clements} mesh.  Small residual dependences on correlation (proportional to $\theta_m \theta_n$) lead to a slight deviation in $\mathcal{E}_0$, while the theoretical curve for $\mathcal{E}_c$ matches very accurately.  The behavior is qualitatively very close to that in the uncorrelated case.

\section{Neural Network Model}
\label{sec:s3}

\begin{figure}[tbp]
\begin{center}
\includegraphics[width=1.00\columnwidth]{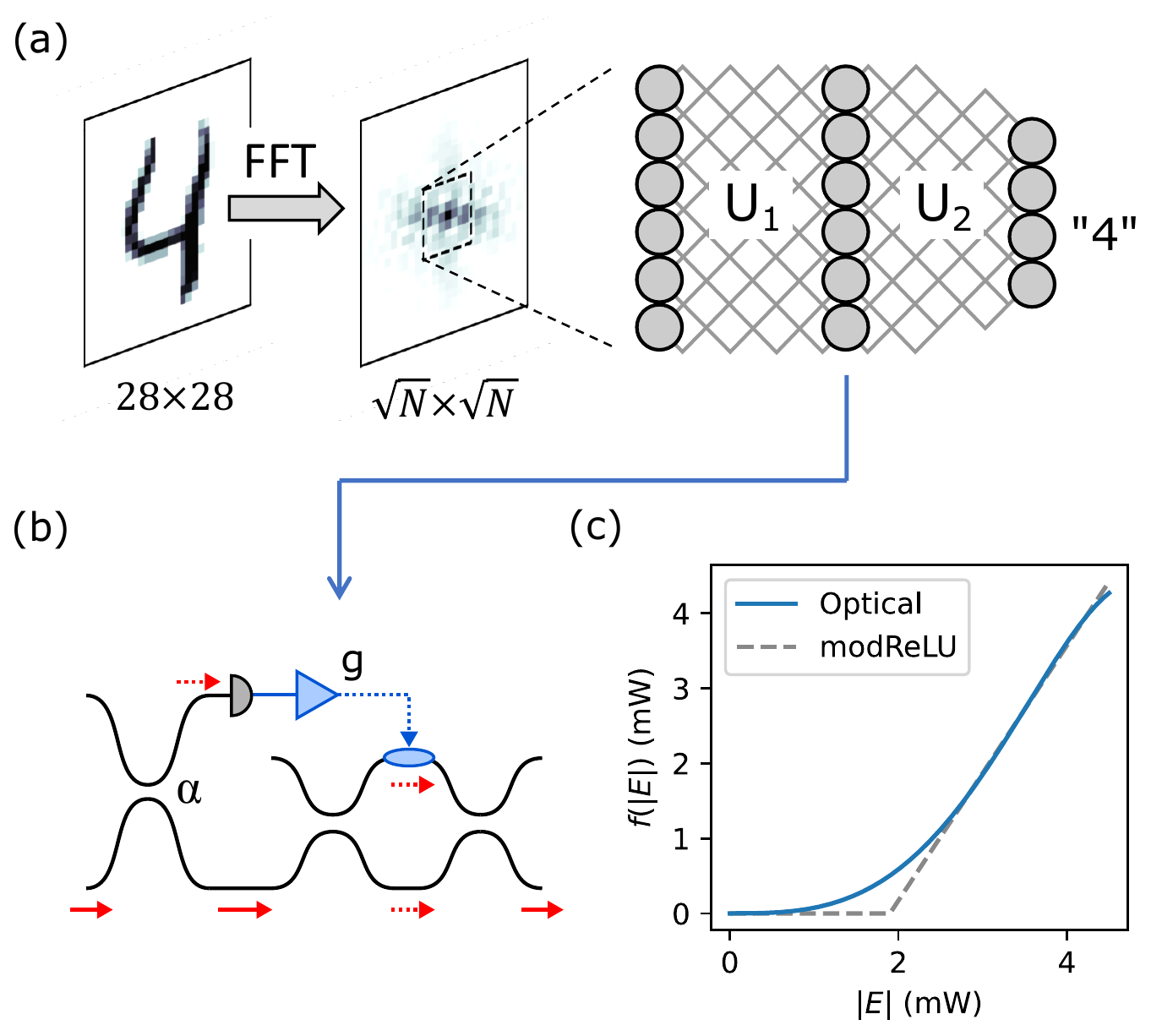}
\caption{Neural network model.  (a) Dataflow: a $28\times 28$ image is Fourier transformed and cropped to an $N\times N$ window; the complex amplitudes are then fed through a two-layer ONN.  (b) Electro-optic nonlinearity implementing Eq.~(\ref{eq:fsupp}).  (c) Input-output relation for the nonlinearity, which approximates a complex modReLU.}
\label{fig:fs4}
\end{center}
\end{figure}

The optical neural network model is based on the architecture described in Ref.~\cite{Pai2020}.  Images from the MNIST digit dataset are preprocessed with a Fourier transform, which is cropped to a $\sqrt{N}\times \sqrt{N}$ window, where $N$ is a model parameter that quantifies the size of the neural network.  The light from this window ($N$ input neurons) is fed into a two-layer optically accelerated DNN.  This DNN consists of a single inner layer and two $N\times N$ unitary circuits, represented by \textsc{Clements} meshes (Fig.~\ref{fig:fs4}(a)).

The activation function at the inner layer is realized with an electro-optic nonlinearity: a fraction of each output field is fed into a detector that drives a Mach-Zehnder modulator, while the remaining light passes through the modulator \cite{Williamson2019}.  This is shown in Fig.~\ref{fig:fs4}(b), and implements the activation function:
\beq
	f(E) = \sqrt{1-\alpha}\, e^{-i(g|E|^2 + \phi - \pi)/2} \cos\bigl(\tfrac{1}{2}(g|E|^2 + \phi)\bigr) \label{eq:fsupp}
\eeq
where $\alpha$ is the power tap fraction, $g$ is the modulator phase induced per unit optical power, and $\phi$ is the phase in the absence of power.  Here, we choose $\alpha = 0.1$, $g = \pi/20$, and $\phi = \pi$, which causes $f(E)$ to approximate the form of a complex modReLU \cite{Arjovsky2016} in the right power regime (Fig.~\ref{fig:fs4}(c)).

Models of sizes $N = 64$ and $N = 256$ were trained using the \textsc{Neurophox} package \cite{Neurophox}.  Code and model parameters are provided in the Supplementary Material \cite{Supp}.

\section{Additional Hardware and Calibration Errors}
\label{sec:s4}

We find that beamsplitter imperfections are the dominant source of error in most situations.  Additional hardware imperfections include: (1) random phase shifts
, which are perfectly correctable provided the error is below a given threshold, and (2) unbalanced loss and finite programming precision, which are in general uncorrectable.  The sections below analyze these effects in detail, concluding that for realistic hardware, errors in the former category are well below the threshold for perfect correction, while errors in the latter category are small enough that they can be ignored.

\subsection{Perfectly Correctable Errors: Phase Shifts
}

While multiport interferometers are length-balanced to ensure that all paths accumulate the same phase shift (in the absence of heater power), in practice there will be phase errors due to localized imperfections.  If these errors are known, they can be calibrated away and absorbed into the programmable phase shifts $(\theta, \phi)$ of each MZI.  Various calibration procedures have been reported in the literature \cite{Suda2017, Alexiev2021}.  However, even without pre-calibration, the method reported in this paper can exactly correct for phase-shifter errors provided that their magnitude falls below a given threshold.

\begin{figure}[tbp]
\begin{center}
\includegraphics[width=1.00\columnwidth]{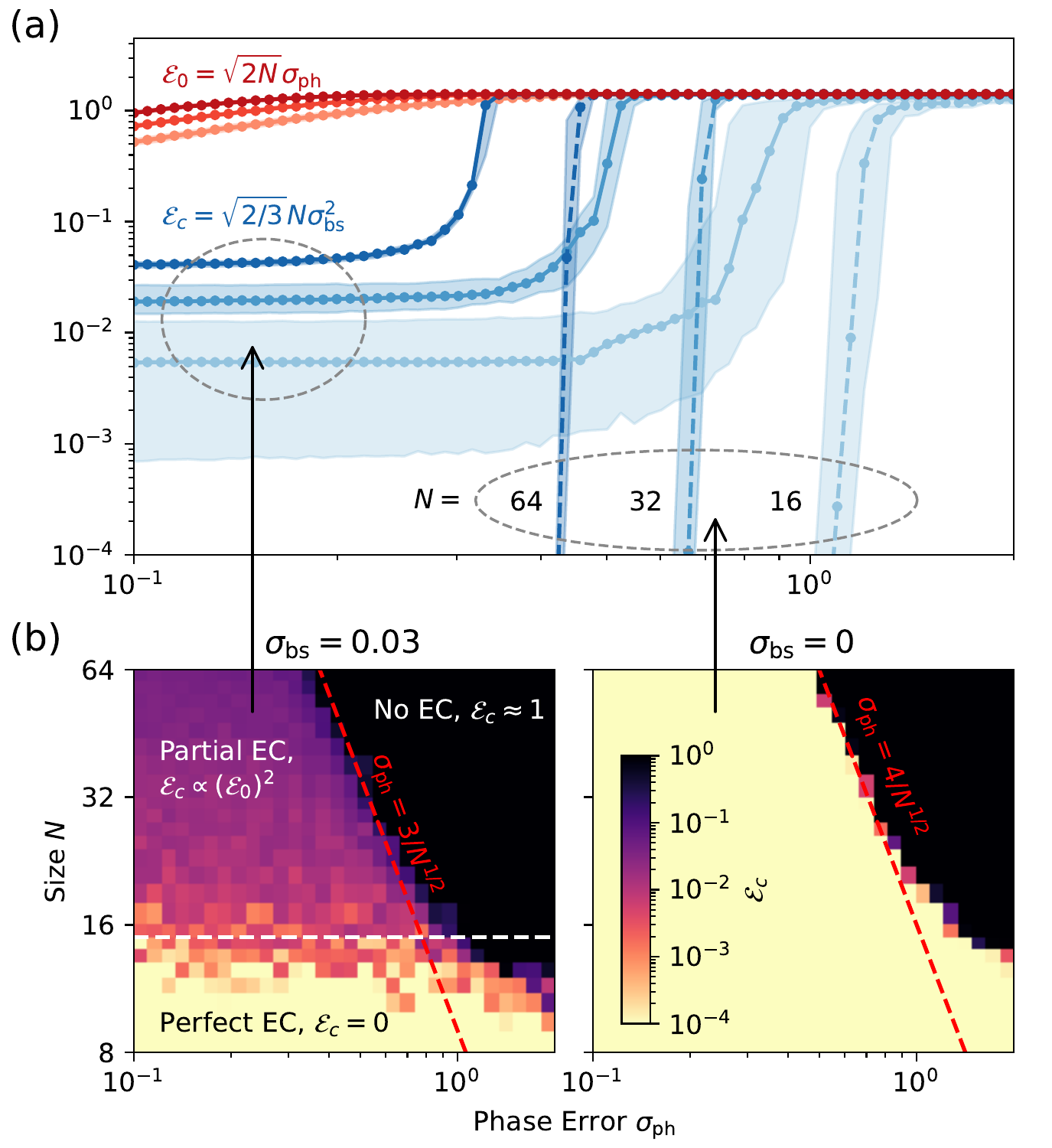}
\caption{Effect of phase errors on self-configuration.  (a) Error as a function of $\sigma_{\rm ph}$ for $\sigma_{\rm bs} \in \{0, 0.03\}$, both uncorrected and corrected.  Note that uncorrected error saturates to unity for $\sigma_{\rm ph} \gtrsim 0.2$.  (b) Corrected error as a function of $N$ and $\sigma_{\rm ph}$, showing the sharp error correction threshold at $\sigma_{\rm ph} \sim 3/N^{1/2}$.}
\label{fig:fs5}
\end{center}
\end{figure}

To start, consider an error model with random phase shifts on both the internal and external arms, i.e.\ a given MZI has the $2\times 2$ unitary
\begin{align}
	& T(\theta, \phi) \rightarrow \nonumber\\
	&\ \ \ \ 	S(\tfrac\pi4 \!+\! \beta) \begin{bmatrix} e^{i(\theta+\psi_{1})}\! & 0 \\ 0 & \!e^{i\psi_{2}} \end{bmatrix}
		S(\tfrac\pi4 \!+\! \alpha) \begin{bmatrix} e^{i(\phi+\psi_{3})}\! & 0 \\ 0 & \!e^{i\psi_{4}} \end{bmatrix}
\end{align}
where errors are uncorrelated and Gaussian: $(\alpha, \beta) \sim N(0, \sigma_{\rm bs})$ and $\psi_i \sim N(0, \sigma_{\rm ph})$.  In Fig.~\ref{fig:fs5}, we numerically perform self-configuration on such faulty \textsc{Clements} meshes, varying both mesh size $N$ and phase error $\sigma_{\rm ph}$, for the cases of ideal ($\sigma_{\rm bs} = 0$) and non-ideal ($\sigma_{\rm bs} = 0.03$) splitters.  In the absence of error correction, the matrix error is calculate to be $\mathcal{E}_0 = \sqrt{2N (\sigma_{\rm ph}^2 + \sigma_{\rm bs}^2)}$, which is dominated by the phase shifts when $\sigma_{\rm ph} > \sigma_{\rm bs}$.  With self-configuration, we see a sharp phase transition between a regime where error correction succeeds ($\mathcal{E}_c = \sqrt{2/3} N\sigma_{\rm bs}^2$) and fails ($\mathcal{E}_c \approx 1$).  This threshold scales as $\sigma_{\rm ph} \sim 3/\sqrt{N}$, and varies only slightly with the splitter error $\sigma_{\rm bs}$.

This threshold arises from imperfect initialization of the mesh, as phase errors cause the initialized MZIs to deviate from the cross state.  While the algorithm does not rely on these MZIs being exact crossings, the measurement steps consist of nulling optical signals that depend on fields propagating down a long chain of cross-like MZIs (dashed lines in Fig.~\ref{fig:f3}).  As the (amplitude) transmission of an imperfect crossing goes as $t_{\rm MZI} \sim e^{-(\psi_1-\psi_2)^2/2}$ for small errors, the average transmission after $N$ stages (there are approximately $N$ crossings during the initial self-configuration steps) is $t = e^{-N \sigma_{\rm ph}^2}$.

The algorithm fails for $\sigma_{\rm ph} > 4/\sqrt{N}$ even with perfect 50:50 splitters because $t < 10^{-7}$, which is below 32-bit machine precision.  In practice, zeroing to such weak signals is impractical in the presence of noise, so realistically one requires $N \sigma_{\rm ph}^2 \lesssim 2$ in order to maintain a reasonable amplitude for the nulling signal $t \gtrsim 0.1$.  This is a factor of 2--3 below the thresholds in Fig.~\ref{fig:fs5}, so self-configuration should work to high accuracy.

In a recent $32\times 32$ SOI switch network, Suzuki et al.\ measured a trimming power variability ($\sigma = 1.52~\text{mW}$) an order of magnitude lower than the switching power ($P_\pi = 18.1~\text{mW}$) \cite{Suzuki2018}.  Putting these together, we calculate $\sigma_{\rm ph} = 0.2$.  This is small enough to comfortably admit self-configuration on meshes up to size $N \approx 64$ ($N \sigma^2 \approx 2.5$).  To configure larger meshes, an initial calibration step \cite{Suda2017, Alexiev2021} will be needed to comfortably reduce the phase errors so that $N \sigma_{\rm ph}^2 \lesssim 2$.  This calibration need only be performed once, as the phase shifts can be saved as the initial ``cross'' state for self-configuring to any matrix $U$.


\subsection{Uncorrectable Errors: Loss and DAC Precision}

\begin{figure}[tbp]
\begin{center}
\includegraphics[width=1.00\columnwidth]{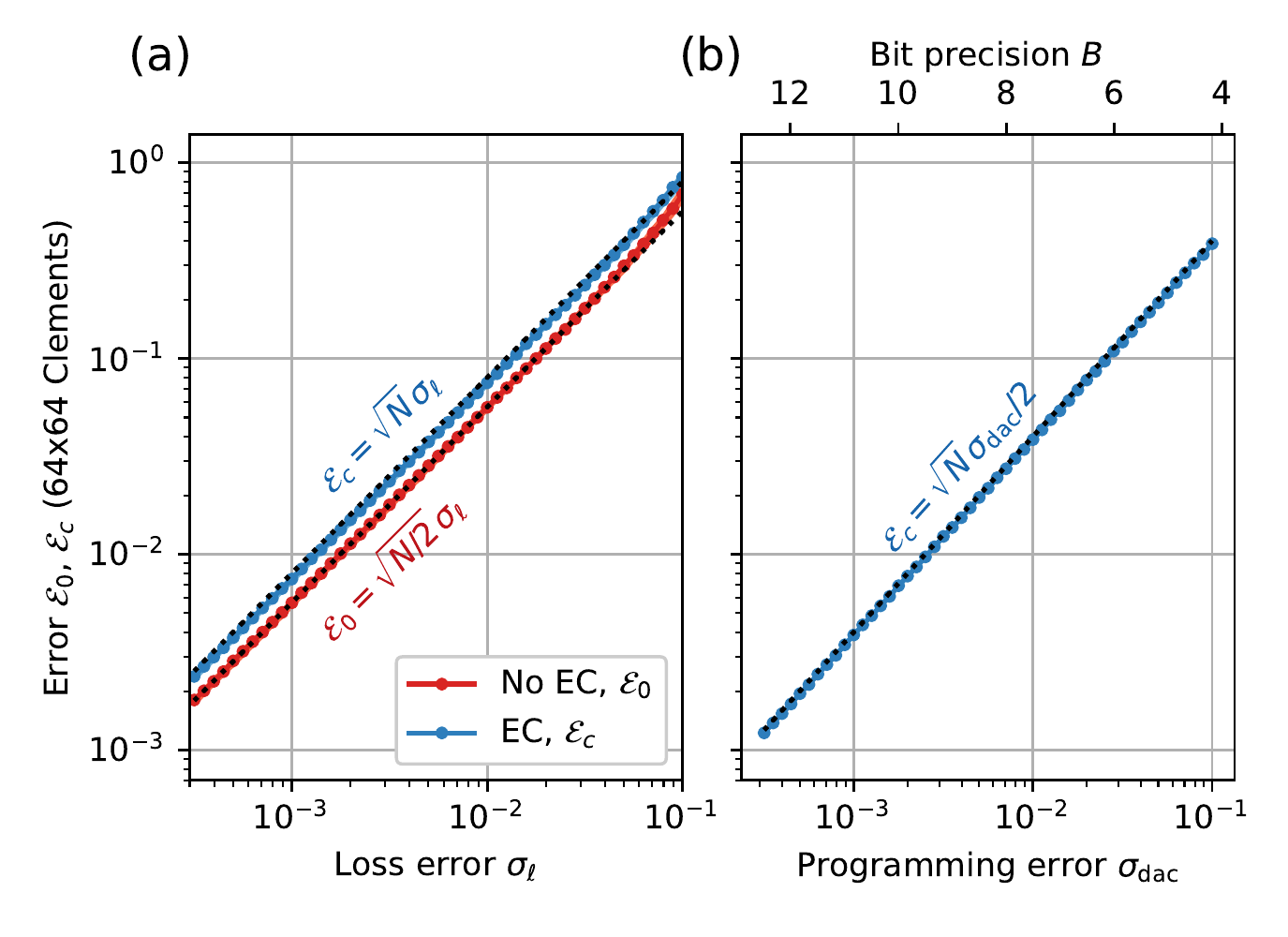}
\caption{Uncorrectable errors, $64\times 64$ \textsc{Clements} mesh.  (a) Effect of loss error on mesh with and without correction.  (b) Programming error due to finite DAC bit precision or noise.}
\label{fig:fs6}
\end{center}
\end{figure}

{\it Loss Errors.}  A uniform waveguide loss does not lead to errors in balanced MZI structures; however, in practice losses can be slightly non-uniform.  We can model loss errors by introducing additional amplitude scalings to the internal and external arms of the MZI, as follows:
\begin{align}
	& T(\theta, \phi) \rightarrow S(\tfrac\pi4) \begin{bmatrix} e^{i\theta-\ell_{1}/2}\! & 0 \\ 0 & \!e^{-\ell_2/2} \end{bmatrix}
		S(\tfrac\pi4) \begin{bmatrix} e^{i\phi-\ell_{3}/2}\! & 0 \\ 0 & \!e^{-\ell_4/2} \end{bmatrix}
\end{align}
where $\ell_i \sim N(0, \sigma_\ell)$ are the loss errors.  The uncorrected error is $\mathcal{E}_0 = \sqrt{N/2}\sigma_\ell$.  We can perform self-configuration on the mesh to attempt to correct the errors, but in practice the error increases by a factor of $\sqrt{2}$ to $\mathcal{E}_c = \sqrt{N}\sigma_\ell$ (Fig.~\ref{fig:fs6}(a)).  The matrix error increases because self-configuration is attempting to correct a non-unitary perturbation with a unitary one, in order to correctly set the splitting ratios of the MZIs.  This procedure is generally successful at setting the MZIs to the target splitting ratios.  However, a nonunitary matrix is not defined solely by its splitting ratio and output phases, so the logic behind error correction breaks down.  In fact, the loss error and the corresponding unitary ``correction'' are orthogonal, so the overall matrix errors add up, leading to the additional factor of $\sqrt{2}$.

Loss errors likely rule out the use of doped-Si thermal phase shifters \cite{Harris2014}, as alignment and doping density variations lead to a significant wafer-scale variation $(0.23\pm 0.13)\,\text{dB}$ ($\sigma_\ell = 0.03$).  However, most SOI platforms use TiN phase shifters, where the heating element is placed sufficiently far above ($\Delta y \gtrsim 0.5~\mu$m) as to not interact with the waveguide mode.  In such phase shifters, loss is determined entirely by the waveguide loss.  Wafer-level statistics for typical SOI processes show $\alpha = (2.1\pm0.25)$~dB/cm for standard processing, while $\text{H}_2$ thermal annealing can reduce this to $(0.1\pm 0.04)$~dB/cm.  Using these figures, for a 200~$\mu$m thermal tuner, $\sigma_\ell = 1.1\times 10^{-3}$ and $1.8\times 10^{-4}$ respectively \cite{Wilmart2020}.  These values are based on wafer-scale variations, so the actual die-scale values relevant to moderate-sized meshes may be much smaller.

Fundamentally, loss variations are limited by the statistics of sidewall scattering.  Consider a waveguide segment of length $L$ with scatterers of size $\approx \lambda/n_{\rm eff}$; if these scatterers are randomly and independently placed, the number in a given segment will follow a Poisson distribution with a mean $\langle n \rangle = L/\lambda$ and a standard deviation $\langle \Delta n \rangle_{\rm rms} = \sqrt{L/\lambda}$.  Dividing these quantities, ratio of loss variation to average loss will be $(\Delta\alpha/\alpha)_L \sim \sqrt{\lambda/L}$.  A more rigorous calculation based on the scattering theory of Lacey and Payne \cite{Lacey1990} (see Sec.~\ref{sec:scatstat}) gives
\beq
	\frac{\langle \Delta\alpha \rangle_{\rm rms}}{\langle\alpha\rangle} = \frac{1}{\pi} \sqrt{\frac{\log(4n_0 L/\lambda)}{n_0 L/\lambda}} \label{eq:avar3}
\eeq
where $n_0$ is the cladding index.  For a 200~$\mu$m phase shifter at 1.55~$\mu$m with $\alpha = 2$~dB/cm, Eq.~(\ref{eq:avar3}) gives $\sigma_\ell = 5\times 10^{-4}$, roughly consistent with the wafer-scale variations in Ref.~\cite{Wilmart2020}.  These loss variations are low enough to allow scaling up to $N \approx 1000$ with matrix errors of at most a few percent.

{\it Programming / DAC Errors.}  The phase shifts $(\theta, \phi)$ are always programmed to finite bit precision; moreover, finite signal-to-noise ratios in the nulling signals may limit the programming accuracy.  To model these effects, in Fig.~\ref{fig:fs6}(b) we self-configure the mesh while introducing a random perturbation to every $\theta$ and $\phi$ after setting it.  For example, if the phase is set by a DAC with bounds $[0, 2\pi]$ and bit precision $B$, truncation to $B$ bits leads to a phase error $\sigma_{\rm dac} = 2\pi/\sqrt{12} \times 2^{-B}$.

Naively, we would expect these errors to add randomly and in quadrature to give $\sqrt{N} \sigma_{\rm dac}$.  However, subsequent steps in self-configuration correct for exactly half of this error.  To see how, note that analogous to Ref.~\cite{SaumilPaper}, for any physical MZI
\beq
	T(\theta, \phi) = S(\tfrac\pi4) \begin{bmatrix} e^{i\theta} & 0 \\ 0 & 1 \end{bmatrix}
		S(\tfrac\pi4) \begin{bmatrix} e^{i\phi} & 0 \\ 0 & 1 \end{bmatrix}
\eeq
there exist $(\psi_1, \psi_2)$ such that $T$ is represented by a symmetric splitter with phase shifts on its outputs:
\beq
	T = \begin{bmatrix} e^{i\psi_1} & 0 \\ 0 & e^{i\psi_2} \end{bmatrix}
		\underbrace{\begin{bmatrix} \cos(\theta/2) & i\sin(\theta/2) \\ i\sin(\theta/2) & \cos(\theta/2) \end{bmatrix}}_{T_{\rm sym}(\theta)}
\eeq
These output phase shifts get corrected in subsequent self-configuration steps.  Therefore, the residual error to matrix $U$ comes entirely from the errors in the $T_{\rm sym}$ blocks
\beq
	\lVert \Delta U \rVert^2 = \sum_{mn} \lVert T_{\rm sym}'(\theta_{mn}) \Delta\theta_{mn} \rVert^2
	= \frac{1}{2} \sum_{mn} \Delta\theta_{mn}^2
\eeq
which averages to $\mathcal{E}_c = \sqrt{N} \sigma_{\rm dac}/2$.  With $B \geq 10$ bits of precision, one can scale to $N \approx 1000$ with matrix errors of at most a few percent.

\subsection{Statistical Origin of Scattering Loss Variations}
\label{sec:scatstat}

Sidewall roughness scattering dominates loss in tightly confined silicon waveguides.  This loss is calculated with the volume-current method, where boundary perturbations lead to current sources that scatter into the far field.  Lacey and Payne show that, in two dimensions, the loss is given by \cite[Eq.~16a]{Lacey1990}:
\beq
	\alpha = C \int_0^\pi \tilde{R}\bigl(\beta - n_0 k_0 \cos\theta\bigr) \d\theta \label{eq:scatt}
\eeq
where $k_0 = \omega/c$, $\beta = n_{\rm eff} k_0$, $n_0$ is the cladding index, and $C$ is a constant of proportionality.  The loss depends on the roughness statistics through $\tilde{R}(k)$, which is the Fourier transform of the autocorrelation function $R(u) = \langle f(z) f(z+u) \rangle$ of the sidewall perturbation $f(z)$.  The integral in Eq.~(\ref{eq:scatt}) corresponds to the sum over the amplitudes of all scatterers $R(\Delta k_z)$ that phase-match the guided mode $k_z = \beta$ to free-space modes $k_z = n_0k_0 \cos\theta$ that lie within the light cone $|k_z| < n_0 k_0$.

While Eq.~(\ref{eq:scatt}) is usually used to calculate the average waveguide loss, sidewall roughness is a statistical quantity and short waveguides will have large relative roughness variations.  We can also use Eq.~(\ref{eq:scatt}) to calculate the statistical loss properties of finite-length waveguides by replacing the loss perturbation with a periodic function:
\beq
	f^{(L)}(z) = \sum_{m>0} \sqrt{\pi \tilde{R}(k_m)} f_m e^{i k_m z} + \text{c.c.}
\eeq
Here $k_m = 2\pi m/L$ is the $m^{\rm th}$ Fourier series mode, while $f_m$ is a complex-valued Gaussian with unit norm in both quadratures, i.e.\ $\langle {\rm Re}[f_m]^2\rangle = \langle {\rm Im}[f_m]^2\rangle = 1$, so that $\langle |f_m|^2 \rangle = 2$.  We calculate the autocorrelation and its spectral function:
\begin{align}
	R^{(L)}(u) & = \frac{2\pi}{L} \sum_m \tilde{R}(k_m) |f_m|^2 \cos(k_m u) \\
	\tilde{R}^{(L)}(k) & = \frac{\pi}{L} \sum_m \tilde{R}(k_m) |f_m|^2 \delta(k - k_m) \label{eq:rlk}
\end{align}
Note that in the limit $L \rightarrow \infty$, $\tilde{R}^{(L)}(k) \rightarrow \tilde{R}(k)$ in the sense of distributions, i.e.\ for any smooth $g(k)$, we have $\int{\tilde{R}^{(L)}(k) g(k) \d k} \rightarrow \int{\tilde{R}(k) g(k) \d k}$, as the former becomes a Riemann sum for the latter and the statistical fluctuations of $|f_m|^2$ average out when many modes are summed.

\begin{table}[tb]
\begin{center}
\begin{tabular}{r|rrrrl}
\hline\hline
	$L = $ & 50 & 100 & 200 & 400 & \ $\mu$m \\ \hline
	$\Delta\alpha/\alpha$ & 0.106 & 0.080 & 0.060 & 0.044 \\
	$\alpha L$ & 0.01 & 0.02 & 0.04 & 0.08 & \ dB \\
	$\Delta \alpha L$ & \ \ 0.0011 & \ \ 0.0016 & \ \ 0.0023 & \ \ 0.0036 & \ dB \\
	$\sigma_\ell$ & 2.5 & 3.7 & 5.5 & 8.2 & \ $\times 10^{-4}$ \\	
\hline\hline
\end{tabular}
\caption{Mean and statistical variation of waveguide loss for a range of lengths, assuming $\alpha = 2$~dB/cm loss.  Ratio $\Delta\alpha/\alpha$ comes from Eq.~(\ref{eq:avar}).  Normalized nonunitary error magnitude $\sigma_\ell$ is given by $\sigma_\ell = \Delta \alpha L/(4.34~\text{dB})$.}
\label{tab:ts1}
\end{center}
\end{table}

We can rewrite Eq.~(\ref{eq:scatt}) as follows:
\beq
	\alpha = C \int_{\beta-n_0k_0}^{\beta+n_0k_0}{\tilde{R}(k) \underbrace{\frac{1}{\sqrt{(n_0k_0)^2 - (\beta - k)^2}}}_{w(\beta-k)} \d k}
\eeq
Following Eq.~(\ref{eq:rlk}), the scattering loss becomes a discrete sum:
\beq
	\alpha = C\frac{\pi}{L} \sum_m {\tilde{R}(k_m) |f_m|^2 w(\beta - k_m)}
\eeq
Note that $\alpha$ depends linearly on the $|f_m|^2$, which are i.i.d.\ random variables.  Applying the moments $\langle |f_m|^2 \rangle = 2$, $\langle (\Delta|f_m|^2)^2 \rangle = 4$, we can calculate the mean and variance of $\alpha$.  These yield discrete sums 
\begin{align}
	\langle \alpha \rangle & = C \frac{\pi}{L} \sum_m {\tilde{R}(k_m) \langle |f_m|^2 \rangle w(k_m)} \label{eq:aav} \\ 
	\langle \Delta\alpha^2 \rangle & = C^2 \frac{\pi^2}{L^2} \sum_m \tilde{R}(k_m)^2 \langle (\Delta|f_m|^2)^2 \rangle w(k_m)^2 \label{eq:daav}
\end{align}
that may be replaced with integrals in the limit of moderately large $L$:
\begin{align}
	\langle \alpha \rangle \rightarrow C \!\int{\!\tilde{R}(k) w(k) \d k},\ \ 
	\langle \Delta\alpha^2 \rangle \rightarrow \frac{2\pi C^2}{L} \!\!\int{\!\tilde{R}(k)^2w(k)^2\d k} \label{eq:daint}
\end{align}
To solve Eqs.~(\ref{eq:daint}), we need to specify roughness model.  For example, a Gaussian model $R(s) = e^{-s^2/2L_c^2}$ would yield $\tilde{R}(k) \sim e^{-k^2 L_c^2/2}$ while an exponential model $R(s) = e^{-|s|/L_c}$ would yield $\tilde{R}(k) \sim 1/(1+k^2 L_c^2)$.  In both cases the model is characterized by a scatterer size $L_c$.  Here we assume that the scatterers are much smaller than the optical wavelength $L_c \ll \lambda$, so that we can approximate the spectral density as a constant $\tilde{R}(k) \rightarrow R_0$, and the model details become irrelevant (in practice $L_c \sim \lambda$ is possible, but the results will be qualitatively similar).  Making this assumption, we find $\langle \alpha \rangle = \pi R_0 C$, while the expression for $\langle \Delta\alpha^2 \rangle$ formally diverges:
\beq
	\langle \Delta\alpha^2 \rangle = \frac{2\pi R_0^2}{L}  \int_{-n_0k_0}^{n_0k_0}{\frac{1}{(n_0k_0)^2 - k^2} \d k} \rightarrow \infty \label{eq:daav2}
\eeq
However, this divergence is merely logarithmic, and the integral is actually just the approximation to a discrete sum Eq.~(\ref{eq:daav}) over a mode set with spacing $2\pi/L$.  Therefore, one should actually impose a cutoff and shrink the bounds of the integral to $\pm (n_0k_0 - \pi/L)$, in which case Eq.~(\ref{eq:daav2}) reduces to

\begin{table*}[t!]
\begin{center}
\begin{tabular}{lc|cc|c|c|ccc}
\hline\hline
& & \multicolumn{2}{c|}{Mesh$^*$} & MVMs & CFLOPs & \multicolumn{3}{c}{Caveats$^\dagger$} \\
\hline
Progressive & \cite{Miller2013} & R & & $N^2$ & $N^3$ & & (D) \\
RELLIM & \cite{Miller2017, Pai2020} & R & C & $N$--$N^2$ & $N^3$ & & (D) \\
In-situ & \cite{Hughes2018} & R & C & $NT$ & $N^3 T$ & & (D) \\
\hline
SGD & \cite{Pai2019} & R & C & $BT$ & $N^2 BT$ & (C) & \\
Numerical & \cite{Mower2015, Burgwal2017} & R & C & $NT$ & $N^3 T$ & (C) & \\
Local & \cite{SaumilPaper} & R & C & -- & $N^2$ & (C) & \\
\hline
Direct & \cite{RyanPaper} & R & & $N^2$ & $N^3$ & & & (S) \\
Ratio & \cite{RyanPaper} & R & & $N^2$ & $N^3$ & & \\
\hline
This work & & R & C & $N^2$ & $N^3$ & & \\
\hline\hline
\end{tabular}
\caption{Comparison of a representative sample of optimization algorithms.  Scaling of computational cost plotted for in-situ resources (number of MVM calls), and in-silico resources (CFLOPs), where $N$ is the mesh size, $T$ is the number of optimization steps, and $B$ is the SGD batch size.  $^*$Mesh types: (R)eck, (C)lements.  $^\dagger$Caveats include (D) requires internal detectors, (C) requires pre-calibration of errors, (S) stability issues in presence of errors.}
\label{tab:ts2}
\end{center}
\end{table*}

\beq
	\langle \Delta\alpha^2 \rangle = \frac{2\pi R_0^2 C^2}{n_0 k_0 L} \log(2n_0k_0 L/\pi)
	= \frac{\lambda R_0^2 C^2}{n_0 L} \log(4n_0 L/\lambda)
\eeq
where $\lambda = 2\pi/k_0$ is the optical wavelength.  Although $\langle \alpha \rangle$ and $\langle \Delta\alpha^2 \rangle$ are functions of model-dependent constants $R_0$ and $C$, their ratio depends only on the refractive index,  wavelength, and length of the waveguide segment:
\beq
	\frac{\langle \Delta\alpha \rangle_{\rm rms}}{\langle\alpha\rangle} = \frac{1}{\pi} \sqrt{\frac{\log(4n_0 L/\lambda)}{n_0 L/\lambda}} \label{eq:avar}
\eeq
Table~\ref{tab:ts1} lists expected losses and their variance for a range of phase shifter lengths and a typical waveguide loss of 2~dB/cm.  The ratio $\Delta\alpha/\alpha \lesssim 0.1$ is also consistent with wafer-scale measurements of waveguide loss variations \cite{Wilmart2020}.  Note that $\sigma_\ell < 10^{-3}$ for reasonable phase shifter lengths, so the expected error $\mathcal{E}_c = \sqrt{N}\sigma_\ell$ is at most a few percent even for meshes as large as $N = 256$.

\section{Algorithmic Efficiency}
\label{sec:s5}

A large number of photonic mesh programming algorithms have been proposed in the literature; see generally Appendix~B of Ref.~\cite{RyanPaper}.  Table~\ref{tab:ts2} summarizes some of the leading approaches in terms of both their computational cost, generality (applicable mesh geometries).  This list is not exhaustive, but gives a representative sample that illustrates the tradeoffs in the field.  With the exception of the algorithms recently reported by us \cite{RyanPaper}, to date all optimization schemes require either (1) accurate pre-calibration of the hardware errors, or (2) $O(N^2)$ internal photodetectors used to monitor power at intermediate points on the mesh.  Our algorithm uniquely lacks both requirements, so it can be performed on uncalibrated ``zero-change'' photonic hardware, requiring only coherent control of the input fields and coherent detection at the outputs, and works for both \textsc{Reck} and \textsc{Clements} meshes.

\begin{figure}[t!]
\begin{center}
\includegraphics[width=1.00\columnwidth]{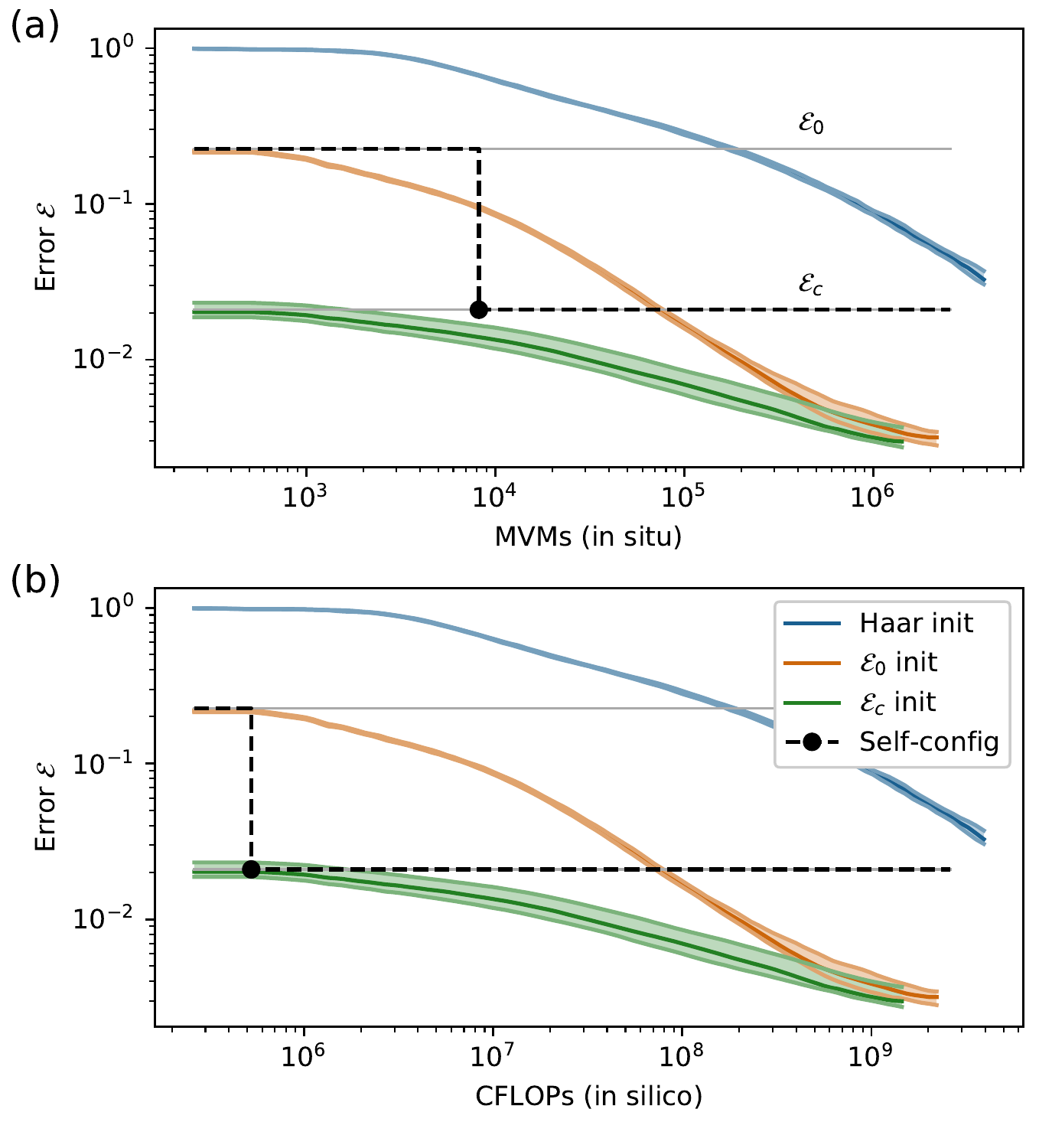}
\caption{Convergence of L-BFGS-B for an imperfect \textsc{Clements} mesh ($64\times 64$, $\sigma = 0.02$), minimizing the $L_2$ norm.  The convergence curve depends on the initial solution, and is compared with the result from self-configuration.  (a) Plot in terms of in-situ MVMs to solution.  (b) In-silico CFLOPs to solution.}
\label{fig:fs7}
\end{center}
\end{figure}

Even in absence of these considerations, our algorithm is competitive against in terms of computational resources.  The required resources of an algorithm depend on whether it is performed in situ (for correction of unknown errors) or in silico (for known, calibrated errors).  For in-situ algorithms, the number of matrix-vector multiplication (MVM) calls is most relevant.  Our algorithm takes approximately $2N^2$ calls (4 calls per MZI), which is comparable to progressive self-configuration and the initial RELLIM proposal \cite{Miller2017, Miller2013} (although RELLIM with internal detectors can be parallelized to run in $O(N)$ time \cite{Pai2020}).  Alternatively, numerical optimization with $T$ time steps takes $3NT$ calls (three calls per step, first to estimate $U$, next to back-propagate $U-\hat{U}$, and finally a forward-pass step to compute the gradient with respect to phase shifts \cite{Hughes2018}).  Fig.~\ref{fig:fs7}(a) shows the convergence of a $64\times 64$ \textsc{Clements} mesh in terms of MVMs under the L-BFGS-B algorithm \cite{Byrd1995}.  Convergence depends strongly on the mesh initialization \cite{Pai2019}.  Initializing to Haar-random unitaries is a significant improvement over random phase shifts, since a mesh with random phase shifts has a banded structure that leads to vanishing gradients with respect to matrix elements far from the diagonal.  Even with Haar initialization, the algorithm takes thousands of steps and millions of MVMs to converge to the accuracy $\mathcal{E}_c$ reached by our algorithm.  Initializing to the phases of an ideal (error-free) \textsc{Clements} mesh helps considerably, but optimization still takes $10\times$ more calls.  However, for \textsc{Clements} meshes it appears that the global minimum can be as much as $5\times$ lower than the self-configuration result; as a result, numerical optimization may still be helpful as a refinement technique, where the mesh is initialized using self-configuration, followed by a numerical procedure to improve this solution.

Since a single optimization step requires $N$ MVM calls, stochastic gradient descent (SGD) has been proposed as a solution to speed up matrix optimization, where a batch of $B < N$ random columns are used rather than the whole matrix.  However, SGD tends to trade off batch size for iteration count, so the overall resource requirement is higher \cite{Pai2019}.

Fig.~\ref{fig:fs7}(b) shows the performance gap with respect to the in-silico case, where accurate calibration of the errors $(\alpha, \beta)$ allows the phases $(\theta, \phi)$ to be computed numerically.  Here, optimization protocols require $O(N^3)$ complex floating point operations (CFLOPs), equivalent to one matrix-matrix multiplication per time step, leading to a scaling of $N^3 T$ (SGD scales as $N^2 B T$).  Self-configuration runs by performing approximately $N^2/2$ Givens rotations with $4N$ CFLOPs each, for a total of $2N^3$ CFLOPs, and thus runs about $10^2\times$ faster.  As before, numerical optimization can still be helpful as a means for further refinement of the solution.

However, for in silico optimization, the local correction method \cite{SaumilPaper} is superior, as it takes only $N^2$ CFLOPs and is parallelizable to $N$ time steps.

\bibliography{PaperRefs}{}
\bibliographystyle{IEEEtranN}

\end{document}